\documentstyle[prl,aps,preprint,psfig]{revtex}
\begin{document}

\title{Magnetic Field Dependence of Electronic Specific Heat in
Pr$_{1.85}$Ce$_{0.15}$CuO$_4$}

\author{Hamza Balci$^a$,V. N. Smolyaninova$^a$,P. Fournier$^b$,
Amlan Biswas$^a$ and R.~L.~Greene$^a$}
\address{a. Center for Superconductivity Research,
Department of Physics, University of Maryland, College~Park,
MD-20742}
\address{b. Centre de Recherche sur les Propri\'et\'es \'electroniques
de mat\'eriaux avanc\'es, D\'epartement de Physique, Universit\'e
de Sherbrooke, Qu\'ebec, CANADA J1K 2R1}
\date{\today}
\maketitle

\begin{abstract}
The specific heat of electron-doped Pr$_{1.85}$Ce$_{0.15}$CuO$_4$
single crystals is reported for the temperature range 2 - 10~K and
magnetic field range 0 - 10~T. A non-linear magnetic field
dependence is observed for the field range 0 - 2~T. Our data
supports a model with lines of nodes in the gap function of these
superconductors. Theoretical calculations of the electronic
specific heat for dirty d-wave, clean d-wave, and s-wave
symmetries are compared to our data.

\end{abstract}

The order parameter (gap) symmetry of the high-Tc cuprate
superconductors(HTSC) is an important parameter in attempting to
understand the pairing mechanism in these materials.  For
hole-doped cuprates experimental evidence strongly favors $d$-wave
symmetry ~\cite{kirtley,harlingen}.  Surprisingly, early
experiments on electron-doped(n-type)
Nd$_{1.85}$Ce$_{0.15}$CuO$_4$ (NCCO) suggested a $s$-wave
symmetry.  Recent penetration depth ~\cite{kokales}, tri-crystal
~\cite{tseui}, photoemission ~\cite{armitage}, Raman
scattering~\cite{blumberg} and point contact tunneling
experiments~\cite{amlan} on NCCO and Pr$_{2-x}$Ce$_{x}$CuO$_4$
(PCCO) favor a $d$-wave symmetry. In addition to these
measurements which show s-wave or d-wave symmetry, there are
penetration depth~\cite{skinta} and point contact
tunnelling~\cite{amlan} experiments that have shown evidence for a
change in the order parameter as the doping changes from
under-doped(d-wave) to over-doped(s-wave). However, since these
prior measurements on the n-type cuprates are surface sensitive
there is a need for bulk measurements (e.g. specific heat) to
convincingly determine the pairing symmetry, as was the case for
the p-type cuprates ~\cite{moler,wright,wang,chiao}.

The specific heat is sensitive to low temperature electronic
excitations. Different gap symmetries have different density of
electronic states close to the Fermi level.  Conventional
low-T$_c$ superconductors show a $s$-wave gap symmetry in which
the electronic specific heat has an exponential temperature
dependence,  $C_{el}\propto T^{1.5}e^{-\Delta/kT}$, where $\Delta$
is the energy gap~\cite{caroli}. For a clean $d$-wave
superconductor electronic excitations  exist even at the lowest
temperatures. The electronic DOS is predicted to have a linear
energy dependence close to Fermi level, and this shows up in the
electronic specific heat as $C_{el}\propto T^2$ ~\cite{volovik}.

In the mixed state, there are two types of quasiparticle
excitations in the bulk of the superconductor: bound states inside
the vortex cores, and extended states outside the vortex cores. In
conventional $s$-wave superconductors, the in-core bound states
dominate the quasiparticle excitations, therefore the electronic
specific heat is proportional to the number of vortices. The
number of vortices is linear in field, therefore the electronic
specific heat is also linear in field. ~\cite{tinkham}. In a
superconductor with lines of nodes(e.g. $d$-wave symmetry), the
extended quasiparticles dominate  the excitation spectrum in the
clean limit. It has been shown that the electronic specific heat
has a $\sqrt{H}$ dependence in the clean limit~\cite{kopnin} at
T=0. For non-zero temperatures there is a minimum field that
depends on temperature after which the $\sqrt{H}$ dependence
should be observed. In the dirty limit the energy scale related to
impurity bandwidth (or impurity scattering rate) is much larger
than the energy scale related to the Doppler shift due to magnetic
field(the dominant mechanism for the clean d-wave case), and much
less than the superconducting gap maximum. In this limit, i.e.
$k_B T<<(H/H_{c2})\Delta_0<<\hbar\gamma_0<<\Delta_0$, where
$\Delta_0$ is the gap maximum and $\hbar\gamma_0$ is the impurity
band width, the magnetic field dependence deviates from
$\sqrt{H}$, and an $H\log(H)$ like dependence is predicted below a
certain field H*, which depends on temperature and impurity
concentration in the sample ~\cite{kubert}.

In this paper we present the first magnetic field dependent
specific heat measurements on n-type cuprates which probe the
symmetry of the superconducting state. The electronic specific
heat has been observed to have a non-linear magnetic field
dependence. The theoretical model for a clean $d$-wave symmetry
fits reasonably well to our data, however there are deviations
from this type of field dependence below $H^*$=0.6 T (Fig.3). We
find that a $H\log(H)$ type dependence gives a better fit to our
data over the whole range, which means our data can better be
described by a dirty d-wave symmetry. It is important to emphasize
that the main point of this work is to address the question of
s-wave vs d-wave, rather than clean d-wave vs dirty d-wave.

The specific heat data was obtained in the temperature range 2 -
10~K and the magnetic field range 0 - 10~T using the relaxation
method~\cite{bachmann}.  The measurements were repeated in two
systems, a home-made setup and a Quantum Design PPMS with some
modifications on the sample holder to remove the field dependence
of the original chip. The addenda consists of a sapphire substrate
with a thermometer and heater, and Wakefield thermal compound to
hold the PCCO crystal. The addenda was measured separately and was
found to have no magnetic field dependence within the resolution
of our experiment($\pm2.5\%$). The experiment was done on several
optimally doped Pr$_{1.85}$Ce$_{0.15}$CuO$_4$ single crystals (the
mass of the crystals was 3-5 mg). The sample heat capacity is
approximately equal to two times that of the addenda at $T=2$ K,
and equal to that of the addenda at $T=10$ K. The crystals were
grown by the directional solidification technique~\cite{peng}. The
samples were characterized with a SQUID magnetometer and found to
be fully superconducting, with similar transition temperatures
T$_c$=22~K$\pm$2~K.

The specific heat of a d-wave superconductor usually has the
following main contributions: the electronic contribution, which
could have the form $\gamma T$ or $\gamma T^2$ depending on the
field and temperature range the measurement is done, the phonon
contribution, which at the temperature range of our experiment can
be written as $\beta T^3$, and a Schottky contribution, which is
caused by spin-1/2 paramagnetic impurities~\cite{gopal}.
Furthermore $\gamma = \gamma(0)+ \gamma(H)$, where $\gamma(H)$
gives the field dependent part of the electronic specific heat
coefficient, and  $\gamma(0)T$ is the residual linear temperature
dependent part of the electronic specific heat. $\gamma(0)$ is
sample dependent, and its origin is not completely understood.
~\cite{moler,wright,wang}. Non-electronic two-level systems away
from the copper-oxide planes are one of the possible candidates
for the origin of this term~\cite{kubert}.  This term has been
found in all hole-doped samples studied~\cite{moler,wright,wang}.

Fig.1 shows temperature dependence of the specific of PCCO heat at
four different fields, 0 T, 1 T, 2 T, and 10 T applied
perpendicular to the ab-plane of the crystal. The field range 0-2
tesla is the relevant field range to extract the gap symmetry
information~\cite{kopnin}, and at H=10 T the sample is completely
in the normal state($H_{c2}=8$T at T=2K). Driving the sample to
the normal state enables us to extract an important parameter,
$\gamma_n=6.7\pm0.5$ mJ/mole-K$^2$, which is needed to compare our
data to theoretical predictions quantitatively. A global fit which
assumes the phonon coefficient, $\beta$, constant for all fields
and $\gamma$ variable gives a $\gamma(0)=1.4\pm0.2$ mJ/mole-K$^2$.
This value of $\gamma(0)$ is consistent with the values found for
$\gamma(0)$ in the hole-doped superconductors($\gamma(0)\approx 1$
mJ/mole-K$^2$ for YBCO~\cite{moler,wright,wang}). The fact that we
do not have any Schottky upturn at low temperatures for any field
shows that our sample is free from a detectable level of magnetic
impurities. From the slope of the lines, obtained through a global
fit, $\beta=0.29\pm0.01$ mJ/mole K$^4$, and a Debye temperature
$\Theta_D=362\pm4$ K has been extracted. These values are in
reasonable agreement with the other published data in the
literature($\beta=0.244$ mJ/mole K$^4$, and
$\Theta_D=382$~K~\cite{khlopkin}).

Since the phonon specific heat is field independent and there is
no Schottky contribution to the specific heat, subtracting the
zero field specific heat from the specific heat at other fields
gives the field dependent part of the electronic specific heat.
Fig.2 shows the field dependent part of the electronic specific
heat,$\gamma(H)$T, vs magnetic field at 3.4~K in the field range 0
- 8~T. Fig.3 shows theoretical fits to the 3.4 K data in the field
range 0 - 2~T. The clean d-wave fit is calculated using the
equation: ~\cite{wang}

\begin{equation}\label{1}
C_{el} = \gamma_n T
\left(\frac{8}{\pi}\right)\left(\frac{H}{H_{c2}/a^2} \right)^{1/2}
for \left(\frac{TH_{c2}^{1/2}}{T_cH^{1/2}}\right)\ll {1}
\end{equation}
where $\gamma_n=6.7$ mJ/mole $K^2$ (from the intercept of our 10 T
data in Fig.~1), $H_{c2}=10$ T, and $a=0.7$(this value was found
experimentally for YBCO~\cite{moler}) are used. The clean d-wave
fit is clearly better than the linear s-wave fit. Possible
non-linear behavior in an s-wave superconductor will be discussed
below.

Even though the clean d-wave fit has a much better consistency
than the linear s-wave fit, there seem to be deviations between
our data and the clean d-wave theory below $H^*=0.6$~T, which
would be expected from a dirty d-wave superconductor. In fact,
being in the dirty limit is not unexpected, since the penetration
depth measurements performed on similar crystals, grown by this
group, were also consistent with dirty d-wave
symmetry~\cite{kokales}. For clean d-wave symmetry the change in
the penetration depth as a function of temperature is linear in
temperature($\Delta$$\lambda$(T) $\propto$ T), whereas a quadratic
temperature dependence($\Delta$$\lambda$(T) $\propto$ T$^2$) is
expected if the nodes are filled by impurity states, i.e. dirty
d-wave. A quadratic temperature dependence has been observed
consistently by two different groups on many crystals they
studied~\cite{kokales}. If a dirty d-wave function of the type
$C_{el}(H)=A \log(B/H)$ is fitted to our data, an excellent fit is
obtained for the fitting parameters $A=6.2\pm0.6$ mJ/mole-K and
$B=17.6\pm4.7$ tesla.  We compared these parameters with the
theoretical predictions~\cite{kubert} calculated from the
equation:

\begin{equation}\label{2}
C_{el}(H) = \gamma_n T \left(\frac{\Delta_0}{8\hbar
\gamma_0}\right)\left(\frac{H}{H_{c2}/a^2}\right)Log\left(\frac{\pi
H_{c2}}{2 a^2 H}\right),
\end{equation}
where $\hbar\gamma_0$ is the impurity band width, $\Delta_0$ is
the superconducting gap maximum, and $a$ is a geometrical factor
related to vortex lattice geometry. Substituting $H_{c2}=10$ T and
$a=1$, the fitting parameter $B= \left(\frac{\pi H_{c2}}{2
a^2}\right)$ can be calculated to be 15.7 tesla, in reasonably
good agreement with the value generated by our data $B=17.6\pm4.7$
tesla.  The other fitting coefficient $A$ can be calculated from
$A=\gamma_n T \left(\frac{\Delta_0}{8
\gamma_0}\right)\left(\frac{1}{H_{c2}/a^2}\right)$. However the
parameter $\gamma_0$, which is related to the density of
impurities in the sample and is sample dependent, is not known.
Therefore by using the experimental $A=6.2\pm0.7$ mJ/mole-K value,
$\hbar\gamma_0$ can be estimated to be 2.1 K, which is in good
agreement with high quality YBCO samples~\cite{taillefer}.

Our analysis at the other temperatures(Fig.4) also produced
results similar to the T=3.4~K data. For T=2.3~K, $A=3.1\pm0.2$
mJ/mole-K and $B=18.0\pm4.7$ tesla, and for T=2.7~K $A=4.3\pm0.5$
mJ/mole-K and $B=18.7\pm4.9$ tesla are found. Theoretically the
coefficient $B$ should be the same for all temperatures, and the
coefficient $A$ should be linearly proportional to the
temperature. The best fits to the data generated the same values
for $B$ within the error range, and the values for $A$ scale with
temperature, even though not in perfect agreement with the theory.

We should mention that the s-wave theory we used to fit our data
neglects non-linear effects that might arise in the vicinity of
$H_{c1}$ due to vortex-vortex interaction or due to a possible
change in the size of the vortex cores.  Some experiments
performed on s-wave superconductors have shown non-linear, even
$\sqrt{H}$, magnetic field dependence for $C_{el}$. However the
field dependence is not consistent for different temperatures,
which means depending on what temperature the field dependence is
probed, the electronic specific heat has a different field
dependence. Different groups have observed $C_{el}\propto H^n$ for
almost any value of n between 0.5 to 1 depending on what material
they studied and at what temperature range they performed their
experiment~\cite{ramirez,sonier}. While we can not definitively
rule out s-wave symmetry as an explanation for our data we believe
that a dirty d-wave symmetry gives the most consistent and
plausible fit to our data.

In conclusion, our specific heat data strongly suggests that the
$d$-wave symmetry in electron doped PCCO at optimal doping is a
bulk property of the material.  However, due to non-magnetic
impurities in our sample, the electronic specific heat follows a
magnetic field dependence of type $H\log(H)$ below H$^*$=0.6 T,
consistent with dirty d-wave symmetry. In addition, the normal
state Sommerfeld constant of PCCO, $\gamma_n$=6.7$\pm$0.4
mJ/mole-K$^2$, has been measured for the first time.

We thank Prof. Peter J. Hirschfeld, and Prof. Chandan Dasgupta for
insightful discussions. P. F. acknowledges the support of the
Canadian Institute of Advanced Research(CIAR), The Canadian
Fondation for Innovation(CFI), The Natural Sciences and
Engineering Research ouncil of Canada(NSERC), The Foundation of
the Universit\'{e} de Sherbrooke. The work at Maryland was
supported by the NSF DMR 01-02350.

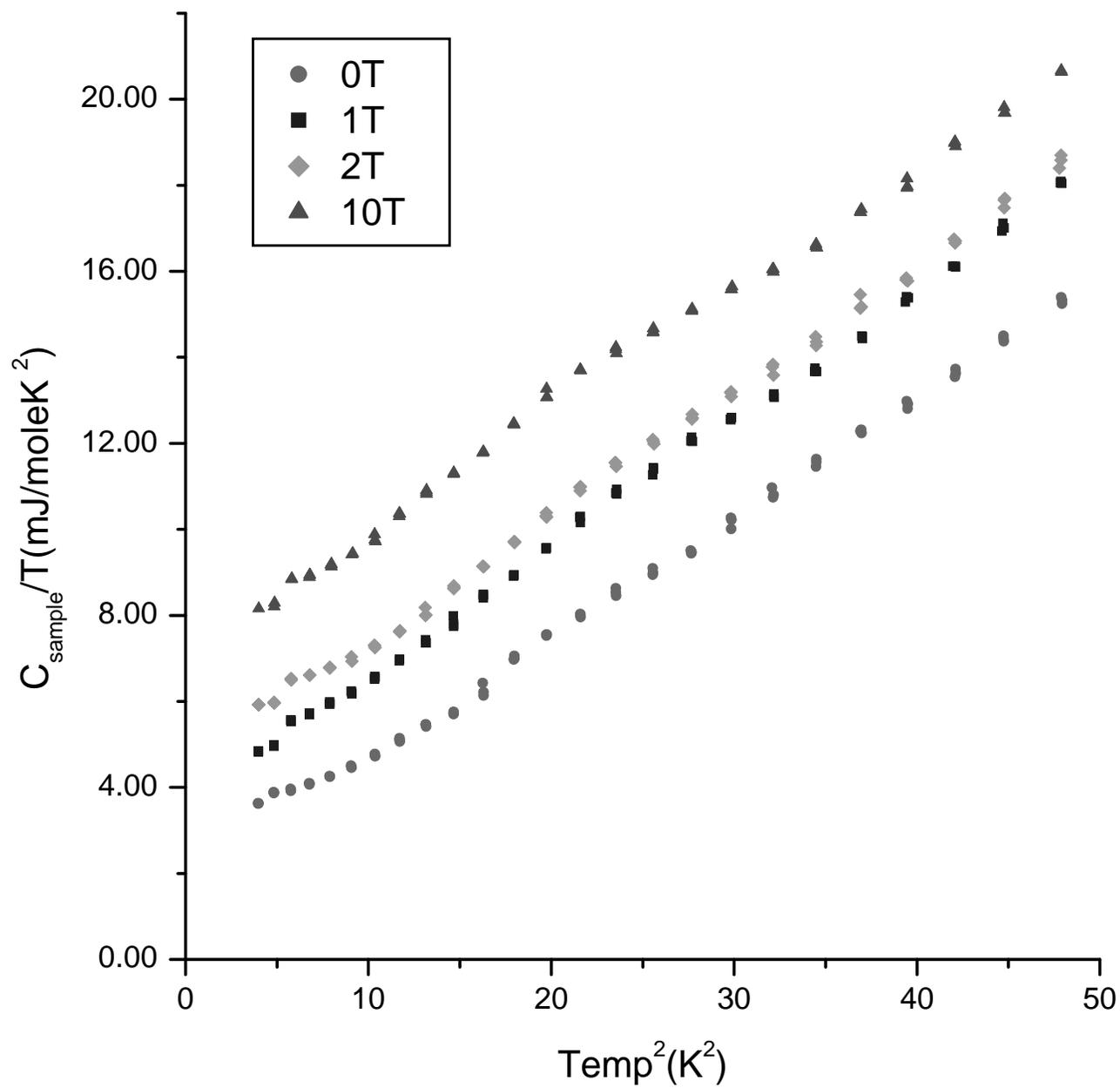
\begin{figure}
\caption{C/T vs T$^2$, where C is the sample total specific heat,
at 4 different fields (0, 1, 2, and 10 Tesla)in the temperature
range 2-7 K. The magnetic field is perpendicular to the ab-plane.}
\end{figure}

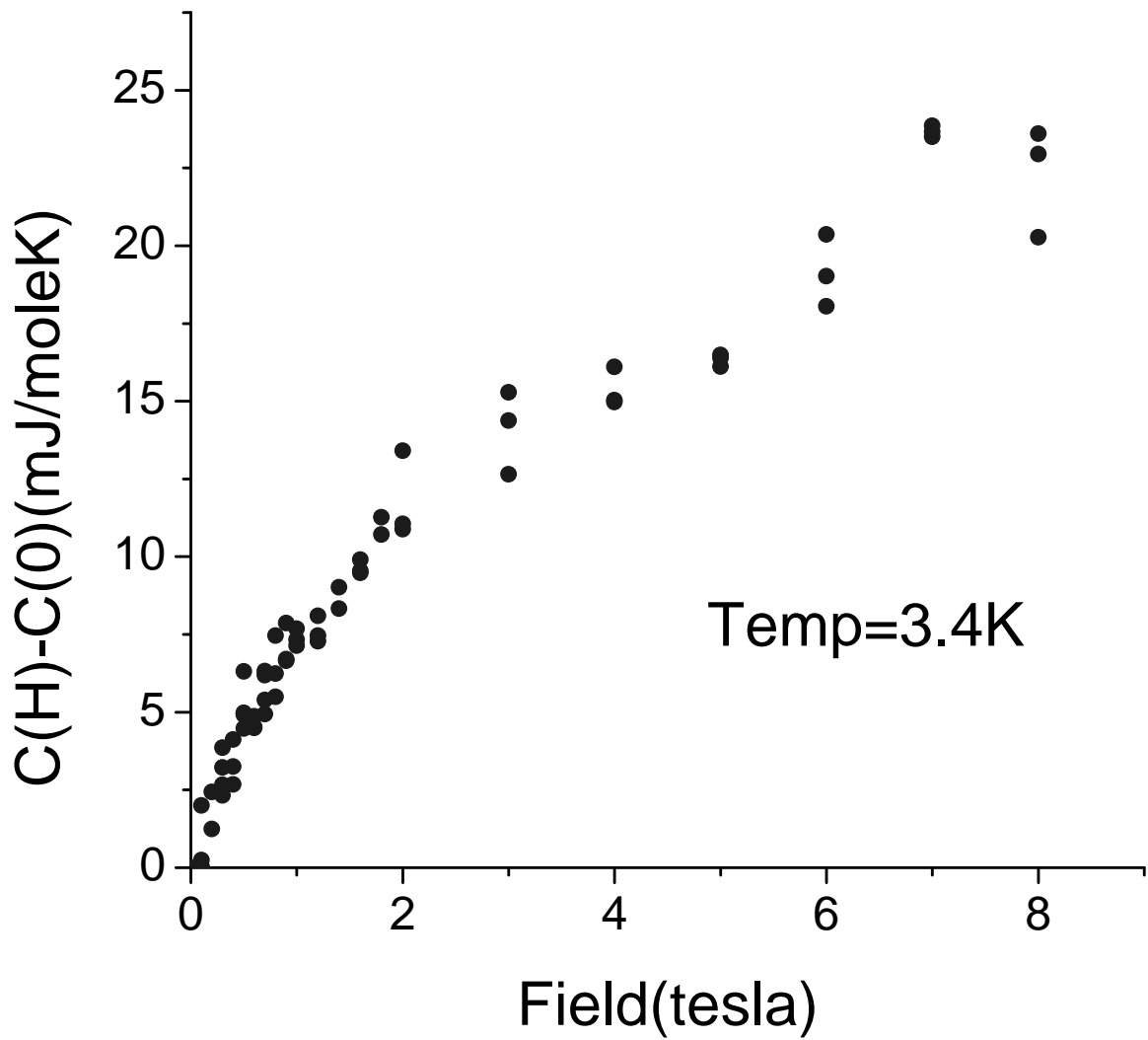
\begin{figure}
\caption{C(H)-C(0) vs magnetic field, or the field dependent
electronic specific heat vs magnetic field. A nonlinear behavior
is observed below 2T, whereas the high field part has a linear
field dependence. The electronic specific heat has almost
saturated to the normal state value at 8~T.}
\end{figure}

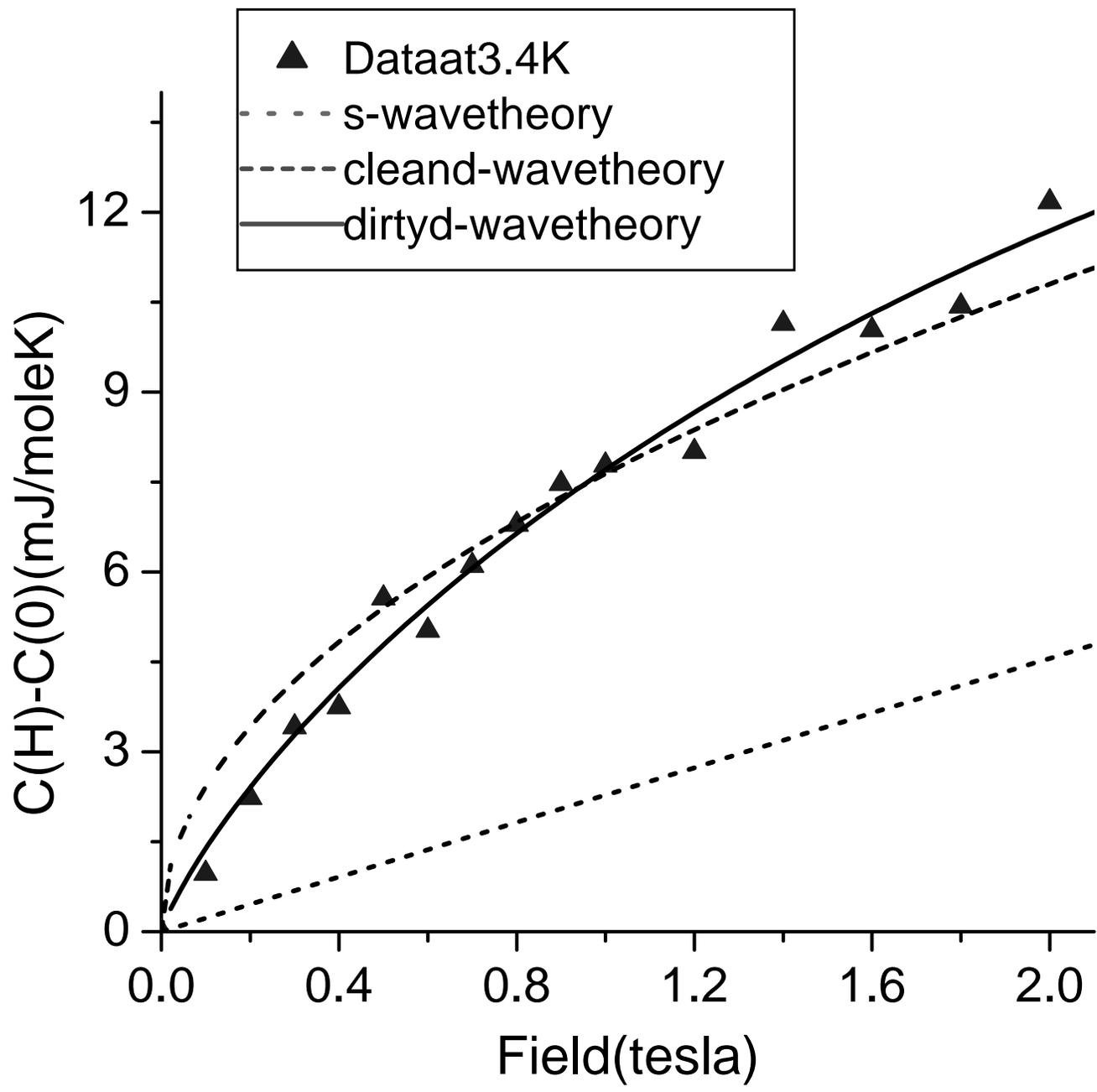
\begin{figure}
\caption{The field dependent electronic specific heat vs magnetic
field data at 3.4 K, and the theoretical fits to the data. The
solid curve is the dirty d-wave fit, the dashed straight line is
the s-wave fit, and the dashed curve is the clean d-wave fit.}
\end{figure}

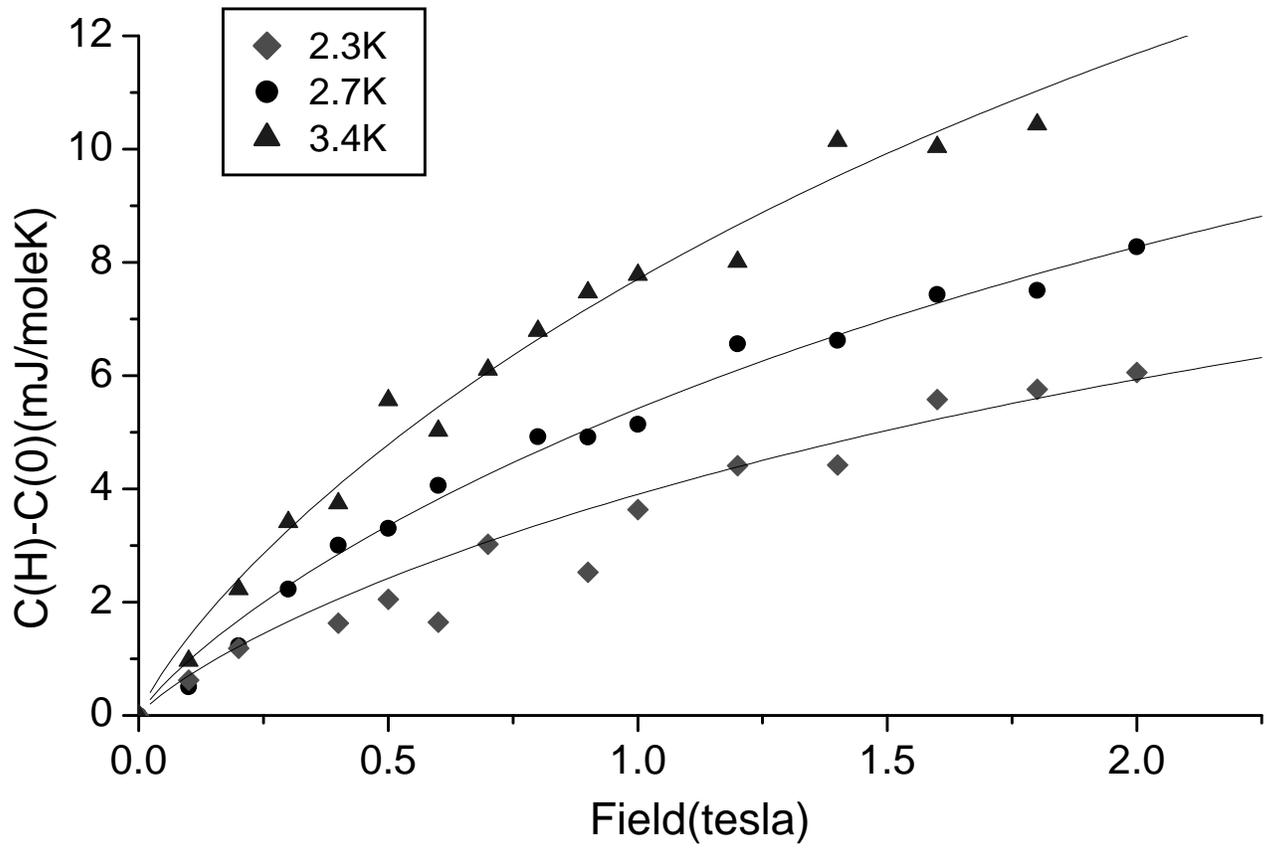
\begin{figure}
\caption{The field dependent electronic specific heat vs magnetic
field data for 2.4K, 2.7K, and 3.4K. The lines are the dirty
d-wave theory fits to the data.}
\end{figure}

\end{document}